\documentclass[conference]{IEEEtran}
    \IEEEoverridecommandlockouts 
    
    \usepackage{cite}
    \usepackage{amsmath,amssymb,amsfonts}
    \usepackage{graphicx}
    \usepackage{xcolor}
    \usepackage{booktabs}
    \usepackage{tabularx}
    \usepackage[hidelinks]{hyperref}
    
\begin{document}
    
    \title{Comparison of sEMG Encoding Accuracy Across Speech Modes Using Articulatory and Phoneme Features}
    
    \author{
    \IEEEauthorblockN{
    Chenqian Le\IEEEauthorrefmark{1},
    Ruisi Li\IEEEauthorrefmark{1},
    Beatrice Fumagalli\IEEEauthorrefmark{1},
    Yasamin Esmaeili\IEEEauthorrefmark{1},
    Xupeng Chen\IEEEauthorrefmark{1},\\
    Amirhossein Khalilian-Gourtani\IEEEauthorrefmark{1},
    Tianyu He\IEEEauthorrefmark{1},
    Adeen Flinker\IEEEauthorrefmark{1},
    Yao Wang\IEEEauthorrefmark{1}
    }
    \IEEEauthorblockA{
    \IEEEauthorrefmark{1}New York University, New York, NY, USA
    }
    \thanks{Chenqian Le and Ruisi Li contributed equally. This work was supported in part by the NYU Discovery Research Fund for Human Health.}
    }
    
    \maketitle
    
    \begin{abstract}
    We test whether Speech Articulatory Coding (SPARC) features can linearly predict surface electromyography (sEMG) envelopes across aloud, mimed, and subvocal speech in twenty-four subjects. Using elastic-net multivariate temporal response function (mTRF) with sentence-level cross-validation, SPARC yields higher prediction accuracy than phoneme one-hot representations on nearly all electrodes and in all speech modes. Aloud and mimed speech perform comparably, and subvocal speech remains above chance, indicating detectable articulatory activity. Variance partitioning shows a substantial unique contribution from SPARC and a minimal unique contribution from phoneme features. mTRF weight patterns reveal anatomically interpretable relationships between electrode sites and articulatory movements that remain consistent across modes. This study focuses on representation/encoding analysis (not end-to-end decoding) and supports SPARC as a robust and interpretable intermediate target for sEMG-based silent-speech modeling.
    \end{abstract}
    
    
    \section{Introduction}
    Silent-speech interfaces (SSIs) aim to restore communication for people with aphasia, anarthria, or other voice disorders. Despite progress from small-vocabulary classifiers to continuous decoding, the accuracy and cross-subject robustness of SSIs remain limited \cite{li2023semg_review,chen2023hdsemg}. Robust biosignal modeling across speech modes is critical for practical SSI deployment in assistive communication settings. In this work, we focus on within-subject encoding analyses across speech modes to inform representational target choices for speech decoding.
    
    A central challenge for sEMG-based SSIs is the choice of the intermediate decoding target space. Discrete phoneme labels are commonly used but are weakly grounded in musculature and may be hard to decode from sEMG signals. Alternatively, articulatory representations explicitly encode the continuous movements of speech articulators (e.g., lips, jaw, tongue) and have a long history of linking acoustics to kinematics \cite{10094711,rebernik2021review}. Building on this, the SPARC (\emph{Speech Articulatory Coding}) framework infers articulatory features from speech audio via acoustic-to-articulatory inversion \cite{cho2024sparc}. SPARC provides 12 kinematic features characterizing lip, jaw, and tongue movement; for voiced speech it also includes pitch and loudness. Because SPARC features directly describe articulatory motions, they may align more naturally with sEMG than phoneme one-hot representations, particularly in silent speech conditions.
    
    To systematically compare articulatory and phonemic representations for sEMG modeling, we adopt analysis tools originally developed for neural signal research. Multivariate temporal response function (mTRF) modeling relates time-lagged stimulus features to neural signals and captures both feature selectivity and temporal dynamics \cite{crosse2016mtrf,DiLiberto2015}. Variance partitioning decomposes explained variance into unique and shared contributions from multiple feature sets \cite{lescroart2015variance,Yu2025}. Beyond end-to-end decoding, encoding analyses provide a principled way to evaluate whether candidate intermediate targets are physiologically aligned with sEMG and robust to mode changes that occur in real SSI use. By modeling time-lagged stimulus--response relationships, mTRF-based encoding enables interpretable comparisons across aloud/mimed/subvocal conditions under matched evaluation protocols. Variance partitioning further distinguishes unique versus shared explanatory power between representations, clarifying whether improvements reflect complementary information rather than redundant correlations. Finally, spatial patterns in encoding weights can inform electrode selection and montage design, which is a practical constraint for wearable SSI deployment. Here, we adapt these tools to the muscle-signal domain to compare the predictive power of articulatory and phonemic representations for modeling sEMG activity.
    
    Specifically, we study \emph{encoding}, i.e., how well stimulus features explain sEMG envelopes, across aloud, mimed, and subvocal speech modes using a forward mTRF model with elastic-net regularization \cite{crosse2016mtrf,zou2005elasticnet}. We fit the model using the open-source \texttt{ADMM\_mTRF} Python implementation \cite{khalilian2025admm_mtrf}. Phone timing is obtained with the Montreal Forced Aligner (MFA) \cite{mcauliffe2017mfa}. Our goal is a controlled \emph{encoding/representation} comparison to inform target-space choices for sEMG-based SSIs, rather than demonstrating end-to-end silent-speech decoding gains. For temporal correspondence, silent sEMG trials are aligned to their paired aloud trials through dynamic time warping (DTW) \cite{sakoe1978dtw}. We decompose explained variance via variance partitioning \cite{lescroart2015variance} to quantify the unique contribution of SPARC vs.\ phoneme features, and analyze encoding weights to assess anatomical specificity. 
    
    \textbf{Contributions.}
    (i) We compare SPARC articulatory features with phoneme one-hot representations for sEMG encoding across three speech modes and show that SPARC yields higher encoding accuracy.
    (ii) We show that aloud and mimed speech achieve comparably high encoding accuracies and that subvocal speech remains significantly above chance.
    (iii) Using variance partitioning, we show that SPARC contributes uniquely to sEMG prediction, whereas phoneme features contribute minimally.
    (iv) We map channels to articulators using mTRF weights, revealing anatomically interpretable structure consistent with electrode placement.
    
    \section{Methods}
    \label{sec:method}
    
    \begin{figure}[t]
    \centering
    \includegraphics[width=\linewidth]{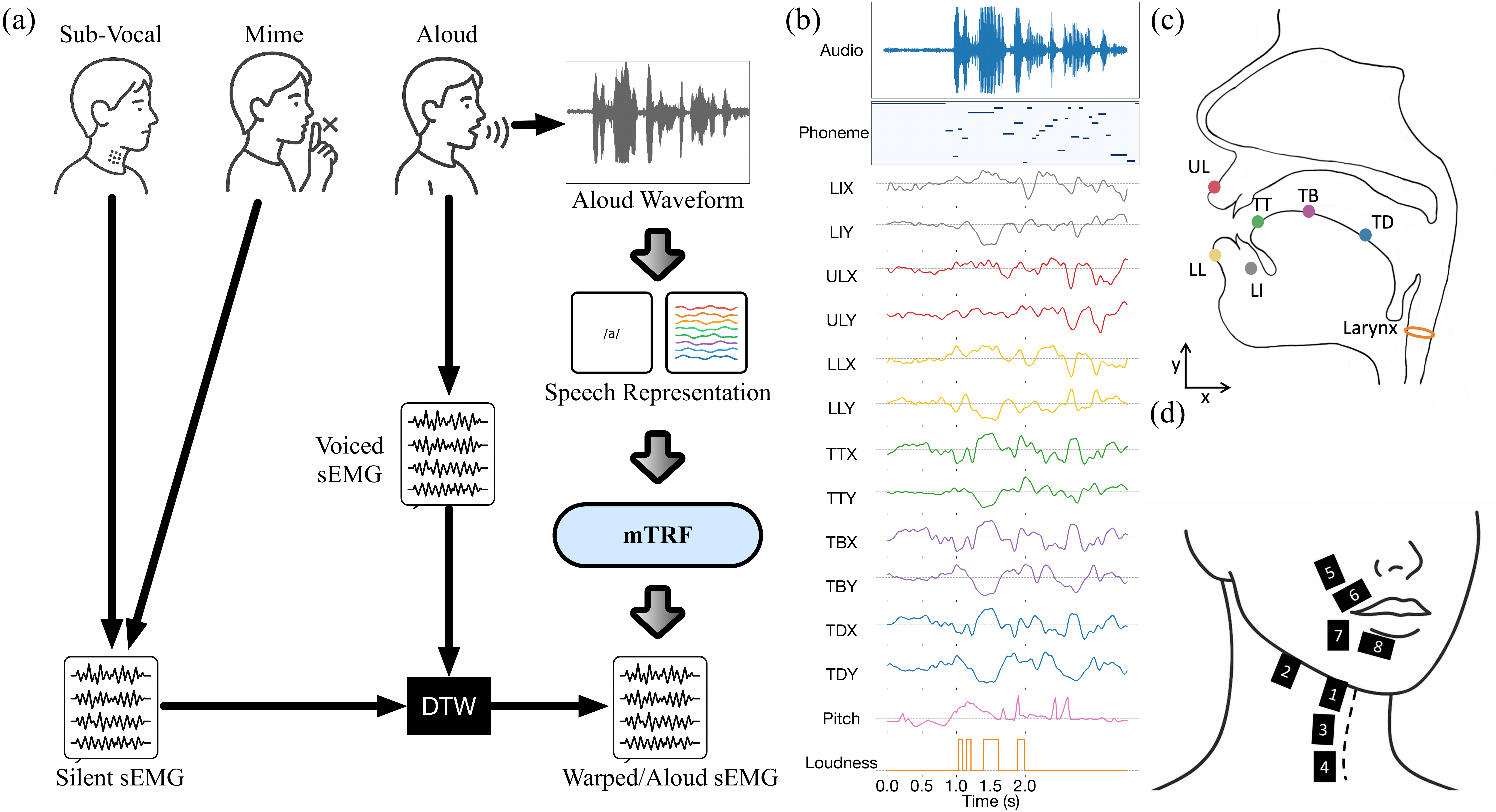}
    \caption{Pipeline and representations. (a) Participants produced each utterance in three modes: aloud, mimed, and subvocal. Eight-channel facial/neck sEMG was recorded. For silent modes, the sEMG envelope was time-aligned to the corresponding aloud envelope using DTW. Representations (SPARC and phoneme sequences) are extracted from aloud audio, and an mTRF predicts the (warped) sEMG. (b) Example representations. (c) SPARC schematic. (d) Eight-electrode sEMG montage.}
    \label{fig:pipeline_dataset}
    \end{figure}
    
    \subsection{Dataset and Preprocessing}
    \label{subsec:data}
    We recruited twenty-four speech-typical participants, each producing $50$ distinct sentences drawn from the TIMIT corpus\cite{garofolo1993timit_ldc93s1}. For every sentence, each participant repeated it three times in each speech mode, resulting in nine utterances per sentence per participant. The speech modes were defined as follows: \emph{aloud} (normal voiced speech), \emph{mimed} (overt articulation without phonation), and \emph{subvocal} (attempted speech with an occluded vocal tract and no phonation).
    
    Surface EMG was recorded from lower face and neck muscles (Ch1--Ch8; Fig.~\ref{fig:pipeline_dataset}(d)). We followed electrode placements as described in \cite{6091201}. The EMG signals were preprocessed using a band-pass filter (10--450\,Hz), together with notch filters at 60\,Hz and its harmonics up to 450\,Hz. Envelopes were extracted as the magnitude of the analytic signal (Hilbert transform) at 2\,kHz and subsequently downsampled to 50\,Hz to match encoding features. To maintain temporal correspondence across modes, each silent trial’s envelope was temporally aligned to its paired aloud envelope using DTW \cite{sakoe1978dtw} and applied to z-scored envelopes (per channel). We used FastDTW with a radius window (radius=30; reduced to 20 only if memory constrained) as the global path constraint, Euclidean distance on per-channel z-normalized envelope vectors as the local cost, and warped the silent envelopes to match the length of the paired aloud reference (silent $\rightarrow$ aloud) for downstream modeling.
    
    \subsection{Speech Representation}
    \label{subsec:speech_representation}
    Audio recordings from aloud speech were band-pass filtered between 50 and 8{,}000\,Hz and downsampled to 16\,kHz. Encoding features for all speech modes were extracted from these downsampled audio signals.
    
    \noindent\textbf{SPARC articulatory features (A).}
    We used \cite{cho2024sparc} to generate SPARC codes from each aloud speech sequence. SPARC comprises 12 continuous kinematic features (upper/lower lip, lower incisor/jaw, tongue tip/blade/dorsum in $x$/$y$) at 50\,Hz. For aloud speech we used the full 14-dimensional SPARC representation, which includes the two laryngeal-related features (pitch and loudness) in addition to the 12 kinematic articulator features; for silent modes we used only the 12 kinematic features, with pitch and loudness excluded rather than set to zero.
    
    \noindent\textbf{Phoneme one-hots (P).}
    We adopted 39 ARPAbet phonemes without stress marks plus one silence token, yielding 40 indices represented as a 40-dimensional one-hot feature. Phoneme labels from MFA were densified to 50\,Hz by repeating labels within each phoneme span. To reduce dominance from leading/trailing silence, we used phoneme timing to retain only speaking intervals of the EMG envelopes and encoding features.
    
    \noindent\textbf{Concatenation (AP).}
    For variance partitioning control analyses, we concatenated features as $[P~A]$.
    
    \subsection{Encoding Model}
    \label{subsec:encoding}
    We denote by $\mathbf{w}\in\mathbb{R}^{F|\mathcal{L}|}$ the concatenated mTRF weights spanning all features and time lags for a given sEMG channel. We initially explored a wider lag window of $[-500,300]$\,ms and found no meaningful structure beyond $\pm300$\,ms; therefore, we used a final window of $[-300,300]$\,ms with 20\,ms steps (matching 50\,Hz sampling), i.e., $\mathcal{L}=\{-300,-280,\ldots,300\}$\,ms. Let $X\in\mathbb{R}^{T\times F}$ denote the feature matrix at 50\,Hz (either $A$, $P$, or $[P~A]$), $y\in\mathbb{R}^{T}$ the envelope for one EMG channel, and $\mathcal{L}$ the set of lags. Let $t=1,\ldots,T$ index time samples at 50\,Hz and $\tau\in\mathcal{L}$ denote a lag (in ms). We form a lagged design matrix
    \begin{equation}
    \mathbf{X}_{\mathrm{lag}}=\bigl[\;X(t-\tau_{1})\;|\;\cdots\;|\;X(t-\tau_{|\mathcal{L}|})\;\bigr]\in\mathbb{R}^{T\times(F|\mathcal{L}|)}.
    \end{equation}
    The mTRF solves a regularized linear regression
    \begin{equation}
    \hat{\mathbf{w}}=\arg\min_{\mathbf{w}}\;\bigl\|y-\mathbf{X}_{\mathrm{lag}}\mathbf{w}\bigr\|_{2}^{2}
    +\alpha\!\left[\,(1-\lambda)\lVert\mathbf{w}\rVert_{2}^{2}+\lambda\lVert\mathbf{w}\rVert_{1}\right],
    \end{equation}
    where $\alpha$ controls regularization strength and $\lambda\in[0,1]$ sets the L1/L2 ratio, and $\mathcal{L}=\{-300,-280,\ldots,300\}\ \text{ms}$. 
    Hyperparameters were selected using an inner cross-validation on the training data only (nested within the outer sentence-level CV). We performed a grid search over $\alpha \in \{10^{-3},10^{-2},10^{-1}\}$ and $\lambda \in \{0.1,0.3,0.5\}$, maximizing average prediction correlation, and the selection consistently favored $\alpha=10^{-2}$ and $\lambda=0.1$, which were then fixed and used for all subjects, channels, and speech modes in the final analyses.
    We used ADMM optimization with a maximum of 10{,}000 iterations, convergence tolerance $10^{-9}$, and penalty parameter $\rho=0.1$.
    
    For each subject, feature set, and speech mode, we computed Pearson correlation $r$ between predicted and observed envelopes on held-out folds, per channel, then averaged across folds (Fisher $z$ for averaging; reported as mean$\pm$SEM across subjects). For paired comparisons we used the Wilcoxon signed-rank test with Benjamini--Hochberg false discovery rate control \cite{benjamini1995fdr,wilcoxon1945}. Chance-level performance was estimated via a permutation test by shuffling contiguous EMG segments indexed by phoneme labels and repeating 1{,}000 times per fold; the 95th percentile of the null distribution was used as the significance threshold. For the permutation test, ``contiguous segments'' are defined as phoneme-labeled spans within each utterance; we shuffle whole spans within an utterance (preserving within-span temporal structure) while keeping the same span boundaries. Segment durations are on the order of $60-200$\,ms (mean $\approx$120\,ms). All evaluations are \emph{within-subject}: sentence-level folds are formed separately for each subject,  as typically done for encoding analysis.
    \subsection{Ethics Statement}
    Approval to carry out the experimental procedures involving human subjects described in this study was obtained from New York University's Institutional Review Board.
    \section{Results}
    
    \subsection{Encoding Across Modes and Electrodes}
\begin{figure}[ht]
    \centering
    \hspace*{-1.5cm}
    \includegraphics[width=1.1\linewidth]{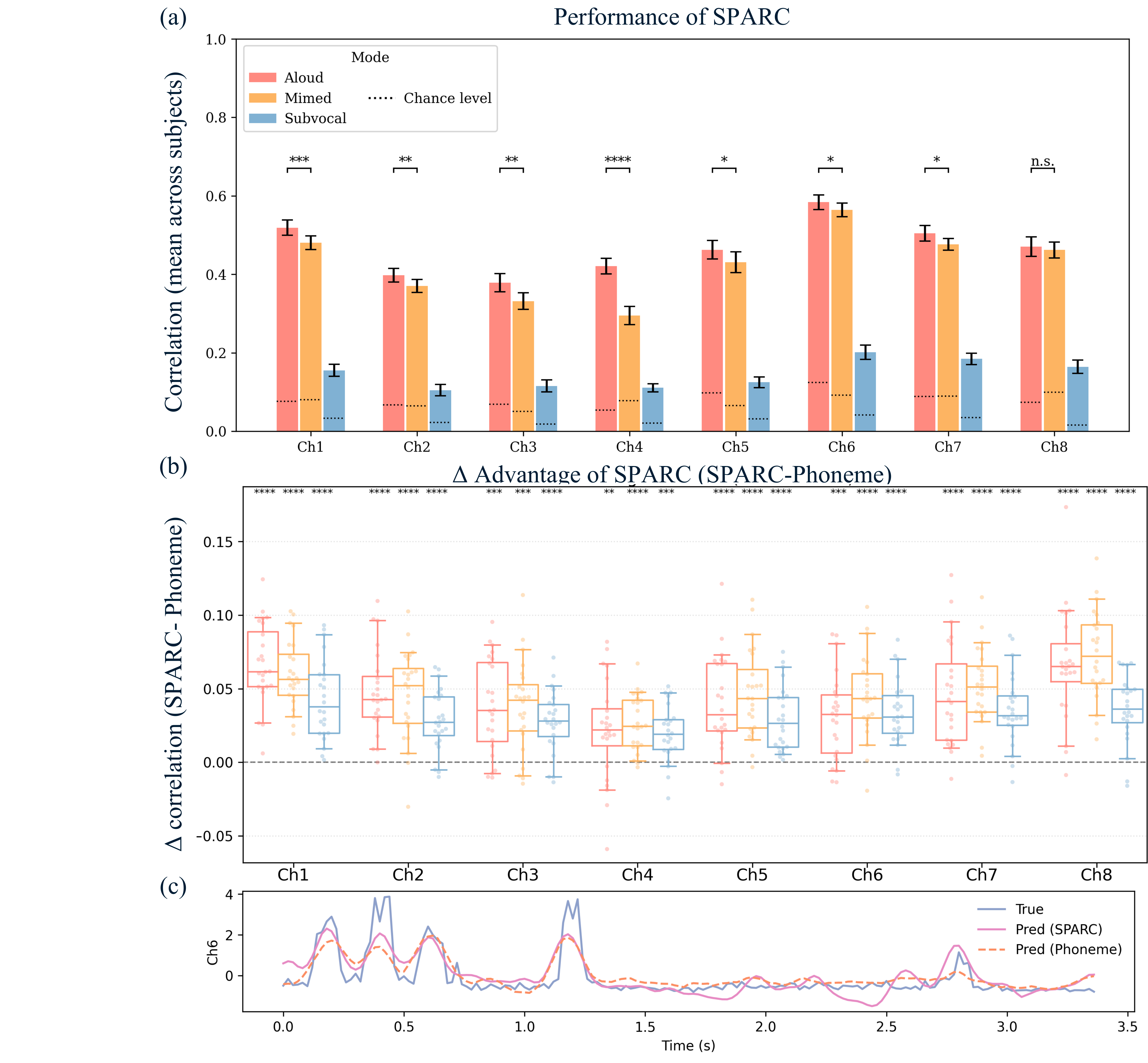}
    \caption{%
            \textbf{Encoding performance by electrode and speech mode.} \textbf{(a)}
            Mean Pearson correlation $r$ (average over 24 subjects) for \textbf{SPARC}
            articulatory features on each channel in \emph{Aloud}, \emph{Mimed}, and
            \emph{Subvocal} speech. Error bars represent standard error of the mean (SEM). Chance level (Dotted black lines) indicates significance threshold by permutation test ($p < 0.05$). \textbf{(b)} Per-subject advantage of SPARC over phoneme one-hots,
            $\Delta r = r_{\text{SPARC}}- r_{\text{Phoneme}}$.
            Stars denote
            Wilcoxon signed-rank tests with FDR control at $\alpha=0.05$ (\textbf{****}\,$p<0.0001$, \textbf{***}\,$p<0.001$, \textbf{**}\,$p<0.01$, \textbf{*}\,$p<0.05$, \textbf{n.s.}\,not significant).  \textbf{(c)} Example of sEMG (Ch 6) predictions from phoneme and SPARC. }
    \label{fig:performance}
    \end{figure}

    We assessed encoding accuracy of SPARC features on each electrode and speech mode. As shown in Fig.~\ref{fig:performance}(a), prediction accuracy (in terms of Pearson correlation) is high for both aloud and mimed speech, with aloud yielding slightly higher accuracy except for Ch4, for which aloud speech had more significant gains. Subvocal speech also yielded performance significantly above chance. The electrode above the upper lip (Ch6) yielded the strongest accuracies in all speech modes. This cross-channel pattern is stable across modes and matches the channel--articulator mapping reported below. 
    
    To quantify the advantage of articulatory features over phoneme labels, Fig.~\ref{fig:performance}(b) plots the per-subject difference $\Delta r = r_{\text{SPARC}}-r_{\text{Phoneme}}$ for each channel and mode. Positive differences are statistically significant on most channels (Wilcoxon signed-rank test with FDR control).
    
    On the Gaddy dataset \cite{gaddy2020digitalvoicing}, SPARC also yielded higher cross-validated encoding accuracy than phoneme one-hots in both voiced and mimed conditions. Averaged across electrodes, Pearson correlation improved from $0.443 \pm 0.017$ (phoneme) to $0.455 \pm 0.021$ (articulatory) in the voiced condition, and from $0.346 \pm 0.029$ to $0.364 \pm 0.032$ in the mimed condition. In each mode, the articulatory model outperformed the phoneme baseline on 7 of 8 electrodes, with statistically significant improvements by a Wilcoxon signed-rank test. Although the mean improvement on the single-subject Gaddy dataset is modest, it is consistent in direction across electrodes and supports the generality of the representational advantage of SPARC under different recording setups.
    
    \begin{figure}[ht]
    \centering
    \includegraphics[width=\linewidth]{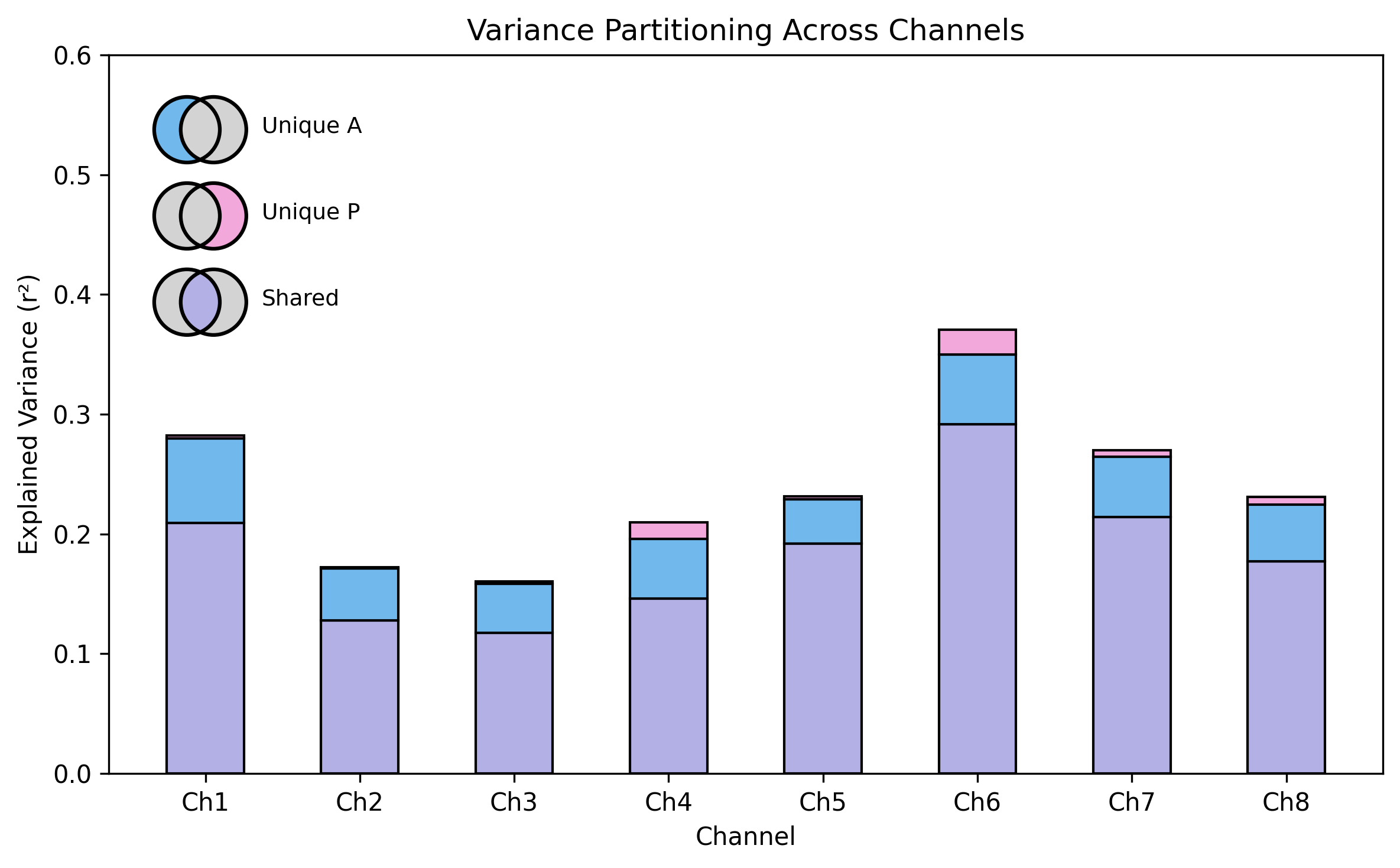}
    \caption{Variance partitioning of explained variance ($r^{2}$) for aloud speech. Bars show unique variance for articulatory features, unique variance for phoneme one-hots, and shared variance, averaged across subjects.}
    \label{fig:variance_partition}
    \end{figure}
    
    We performed cross-validated variance partitioning \cite{lescroart2015variance} using three models: articulatory-only ($A$), phoneme-only ($P$), and concatenation ($AP$). Denoting explained variances ($r^2$) by different features as $r^2_{A}$, $r^2_{P}$, and $r^2_{AP}$, we compute
    \begin{align}
    r^{2}_{\text{Unique A}} & = r^{2}_{AP}-r^{2}_{P}, \\
    r^{2}_{\text{Unique P}} & = r^{2}_{AP}-r^{2}_{A}, \\
    r^{2}_{\text{Shared}}   & = r^{2}_{A}+ r^{2}_{P}- r^{2}_{AP}.
    \end{align}
    As illustrated in Fig.~\ref{fig:variance_partition}, the shared component dominates across electrodes, while articulatory features contribute unique variance substantially larger than phoneme features for all channels.
    
    The comparable performance between aloud and mimed speech suggests that mimed speech still induces measurable  articulatory muscle activities. Lower subvocal performance is consistent with reduced articulatory amplitude and lower signal-to-noise ratio (SNR) under occluded, non-phonated production.
    
    \subsection{Anatomical Specificity}
    \begin{figure}[ht]
    \centering
    \includegraphics[width=\linewidth]{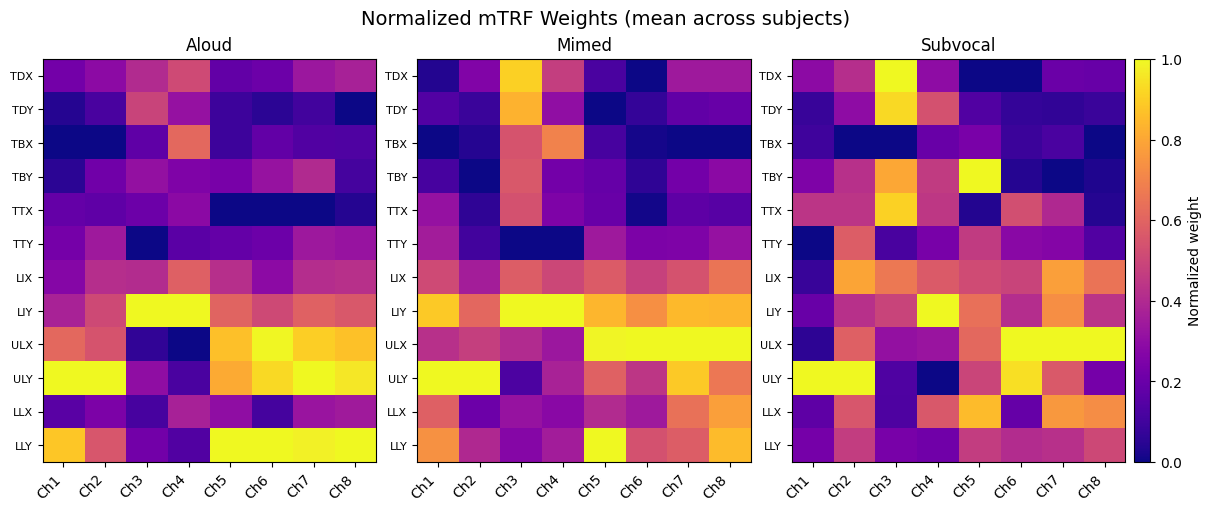}
    \caption{Per-channel normalized linear weight maps (mean across subjects) for aloud, mimed, and subvocal speech. Rows correspond to SPARC features and columns to sEMG channels.}
    \label{fig:weight_maps}
    \end{figure}
    
    Fig.~\ref{fig:weight_maps} summarizes the sum of absolute weights across time lags for each articulatory feature and channel, averaged across subjects and normalized within channel. Peri-oral channels (Ch5--Ch8) are primarily influenced by lip movements across modes. Submental channels (Ch1--Ch2) capture mostly lip movement with secondary jaw contribution. Laryngeal and upper-neck channels (Ch3--Ch4) reflect both jaw and tongue movements, with increased tongue contribution in silent modes. These patterns align with expected biomechanics and remain stable across modes.
    
    While DTW can inflate absolute correlations by smoothing timing differences, it is required here because mTRF regression assumes predictors and responses share a common time index. Importantly, within each trial the same warping path is applied regardless of whether predictors are SPARC or phoneme one-hots, so DTW alone cannot trivially explain a \emph{relative} advantage of SPARC over phoneme features.
    
    Across subjects, the weight patterns show consistent lip/jaw/tongue-dominant channel structure, suggesting stable biomechanical coupling between electrode locations and articulator movements; a more formal quantitative anatomy-consistency test is left for future work.
    
    \section{Conclusion}
    Across twenty-four subjects and three speech modes, SPARC consistently outperformed phoneme features for sEMG encoding, and the same trend was observed on the single-subject Gaddy dataset. Variance partitioning showed substantially larger unique SPARC contributions than unique phoneme contributions, alongside a dominant shared component. Weight maps revealed interpretable relationships between measured muscle activity and articulatory movements. These findings support articulatory targets as a physiologically grounded and interpretable choice for robust sEMG-based SSI modeling, and the encoding weights provide practical cues for electrode placement. Future work will evaluate whether and to what extent these representational advantages translate to downstream decoding performance in end-to-end SSI pipelines.
    
    \bibliographystyle{IEEEtran}
    \bibliography{strings,refs}
    
    \end{document}